\documentclass[useAMS,usenatbib]{mn2e} 
\usepackage{amsmath}
\usepackage{amssymb}
\usepackage{float}
\usepackage{graphicx}
\usepackage{lscape}
\usepackage{natbib}

\title[The Creation of AGB Fallback Shells]{The Creation of AGB Fallback Shells}
\author[Z. Chen, A.
 Frank, E. G. Blackman \& J. Nordhaus]
{Zhuo Chen$^{1}$\thanks{E-mail:
zchen25@ur.rochester.edu},
Adam Frank$^{1}$\thanks{E-mail: afrank@pas.rochester.edu },
Eric G. Blackman$^{1,2}$\thanks{E-mail: blackman@pas.rochester.edu} and
Jason Nordhaus$^{3,4}$\thanks{E-mail: nordhaus@astro.rit.edu} \\
$^{1}$Department of Physics and Astronomy, University of Rochester, Rochester NY, 14627\\
$^{2}$School of Natural Sciences, Einstein Drive, Institute for Advanced Study, Princeton NJ, 08540\\
$^{3}$National Technical Institute for the Deaf, Rochester Institute of Technology, 52 Lomb Memorial Drive, Rochester NY, 14623\\
$^{4}$Center for Computational Relativity and Gravitation, Rochester Institute of Technology, One Lomb Memorial Drive, Rochester NY, 14623}
\begin{document}

\date{Written in 2015 April 30}

\pagerange{\pageref{firstpage}--\pageref{lastpage}} \pubyear{2015}

\maketitle

\label{firstpage}

\begin{abstract}

The possibility that mass ejected during Asymptotic Giant Branch (AGB) stellar evolution phases falls back towards the star has been suggested in applications ranging from the formation of accretion disks to the powering of late-thermal pulses.  In this paper, we seek to explicate the properties of fallback flow trajectories from mass-loss events.  We focus on a transient phase of mass ejection with sub-escape speeds,  followed by a phase of a typical AGB wind.  We solve the problem using both hydrodynamic simulations and a simplified one-dimensional analytic model that matches the simulations. For a given set of initial wind characteristics, we find  a critical shell velocity that distinguishes between "shell fallback" and "shell escape". We discuss the relevance of our results for both single and binary AGB stars.  In particular, we discuss how our results help to frame further studies of fallback as a mechanism for forming the substantial population of observed post-AGB stars with dusty disks.

\end{abstract}

\begin{keywords}
stars: AGB and post-AGB --- stars: winds, outflows --- methods: numerical
\end{keywords}

\section{Introduction}

The possibility that some material ejected during the AGB mass loss stage falls back onto the star has been the subject of a number of studies, motivated by both theoretical and observational considerations \citep{bujar1998,soker2001,van1998,van1999,waters1992,zijlstra2001}.

The discovery of a population of post-AGB stars hosting dusty, long-lived disks is a puzzle for  standard stellar evolution theory as it is not clear how these disks originate \citep{clayton2014,su2007,venn2014}. Given that some occur in binaries, it has been assumed that most if not all, are the result of some form of mass transfer \citep{bujar2005,dermine2012,hinkle2007,malek2014,van2009}.

Theoretically, there is diverse interest in AGB mass-loss fallback.  Asking when the return of matter can significantly influence post-AGB evolution, \cite{soker2001} considered the forces acting on back-flowing material in order to set limits on when such fallback can occur (though no trajectories were calculated).  In order for back flow to play a role in post-AGB and Planetary Nebulae (PN) evolution, that study concluded that the fallback mass should be high, the ejected velocity should be low, and that the mass should be concentrated along the AGB equator.

For a spherically symmetric case, thermal pulses or fuel limited relaxation oscillations may be a source of such mass loss \citep{vanhorn2003}. In \cite{frankowski2009}, backflows were invoked as a means of creating very late thermal pulses (VLTP).  \cite{hajduk2007} also considered the possibility that mass accretion can induce a VLTP in the context of the ‘old nova’ CK Vul where the mass is assumed to come from a companion. \cite{frankowski2009}, however, considered the accreted mass falling back from AGB wind material that had become part of a planetary nebula.

In \cite{soker2008}, the possibility for fallback to occur via a stellar "evanescent zone" was explored where, in addition to the escaping wind, gas parcels do not reach the escape velocity but rise slowly through the zone and then fall back.  Wind and bound gas were found to exist simultaneously out to distances of $\sim 100$ AU.

Backflow has also been proposed as a mechanism for slowing evolution along post-AGB evolutionary tracks \citep{zijlstra2001} and for depleting post-AGB dust stars of refractory elements (e.g. \cite{van1998}).

Explaining the observations of disks around post-AGB stars (discussed above) has, however, been one of the main reasons for exploring studies of fallback during AGB mass loss.  Such a suggestion was part of the motivation driving the \cite{soker2001} paper.  In addition, \cite{akashi2007} suggested that circumbinary rotating disks can also be formed from slow-AGB-wind material that is pushed back towards the star by wide jets emanating from one of the stars (i.e. collimated PPN or PN outflows \cite{bujar1998}).

A natural means for generating fallback in the context of binary stars however may be common envelope (CE) evolution.  Both simulations and analytic models indicate that some fraction of the material expelled from the envelope may not escape the binary system \citep{nordhaus2006,nordhaus2007,kashi2011,ricker2012,passy2012}.  Some of this material is, therefore, expected to fall back after the CE mass ejection event. This can occur even when the CE event involves a low-mass object such as a massive planet being swallowed by an AGB star\citep{nordhaus2006}.  Planetary companions to main-sequence stars are plentiful and those orbiting within $\sim10$ AU can be expected to plunge into their host stars during the giant branches \citep{nordhaus2010,nordhaus2011,nordhaus2013}.  In such cases, material would be ejected over short timescales (a shell) with sub-escape velocities ($V_s <  V_{esc}$).  Thus, it is possible that circumbinary disks are the result of CE ejections in which some fraction of the envelope is ejected at sub-escape speeds which may drop back to be shaped into a disk.

Given the interest in fallback on the AGB, in this paper we attempt to gain some insight into the dynamics of fallback shells by considering only the radial trajectories of the expelled gas.  Of particular interest is the response of a sub-escape velocity shell to internal driving by the AGB wind which would likely follow the shell ejection.  Note that we do not explicitly specify or model the origin of the mass-loss history as our goal here is simply to study the dynamics of the shell after its initial ejection from the star.  In Sect. 2, we introduce our hydrodynamic model and methods to solve the problem. In Sect. 3.1, we present the results of full hydrodynamic simulations.  In Sect. 3.2 we present a semi-analytic model and compare it with the simulation results as well as present a zeroth order estimate of the critical velocity of the shell that distinguish collapse or escape trajectories.

\section{Method and Model}

We use the \textsc{AstroBEAR} adaptive mesh refinement(AMR) code to solve the hydrodynamic equations \citep{cunningham2009,carrol2013}. AMR is a computational technique that divides a computational cell into $2^{dimension}$ child cells when the physical condition suggests to (usually high gradient).

\begin{equation}\label{mass eq}
\frac{\partial \rho}{\partial t}+\nabla \cdot(\rho \mathbf{v})=0
\end{equation}
\begin{equation}\label{momentum eq}
\frac{\partial \rho \mathbf{v}}{\partial t}+\nabla \cdot (\rho \mathbf{v} \mathbf{v})=-\nabla P-\frac{G M \rho}{r^2}\mathbf{\hat{r}}
\end{equation}
\begin{equation}\label{idealgas}
P=n k_b T
\end{equation}
where $n=\rho/m_{mean}$ is the particle number density and $m_{mean}=1.3\ m_H$ is the mean particle mass.

\begin{figure}
    \centering
    \includegraphics[scale=0.4]{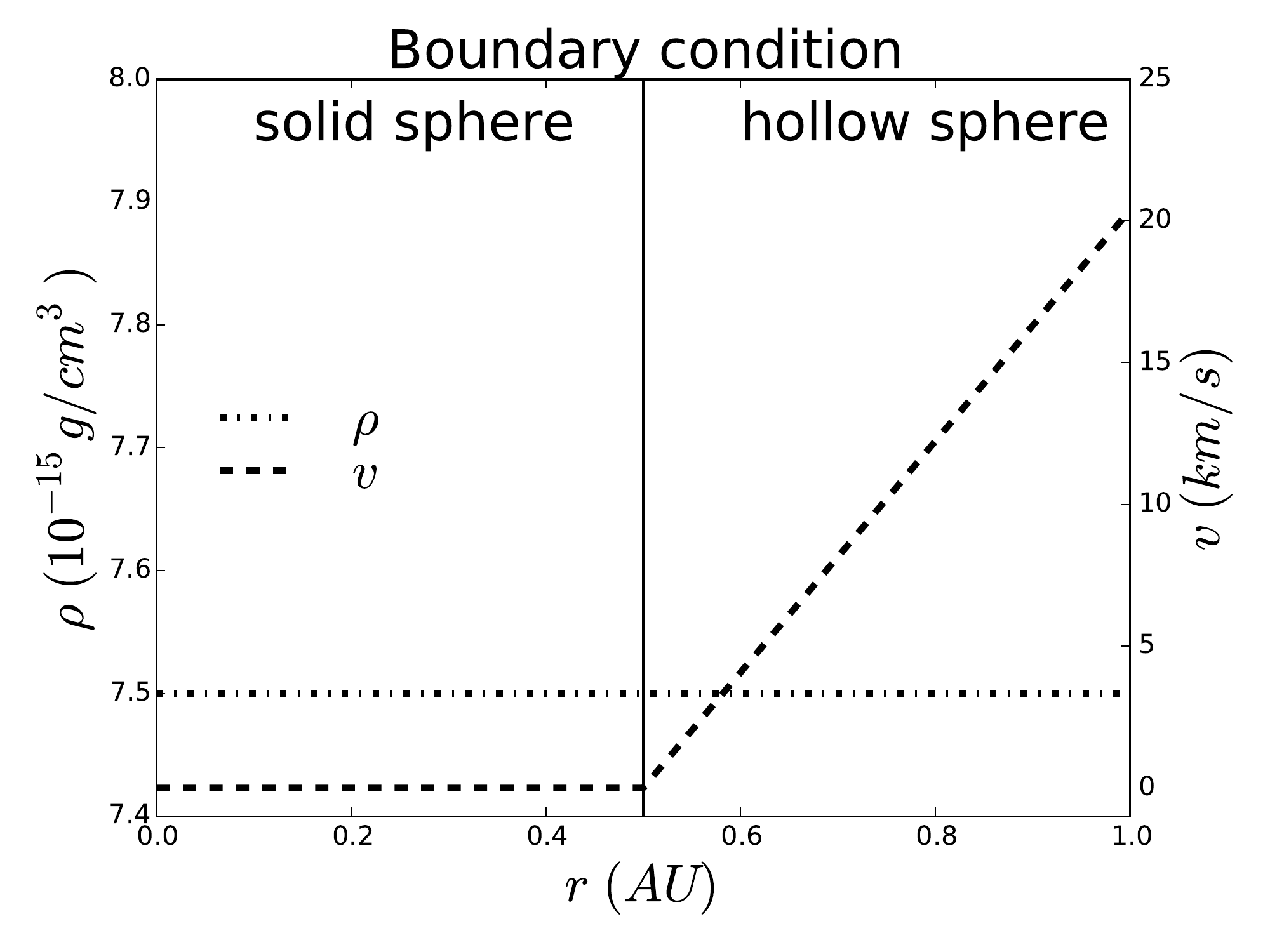}
    \includegraphics[scale=0.45]{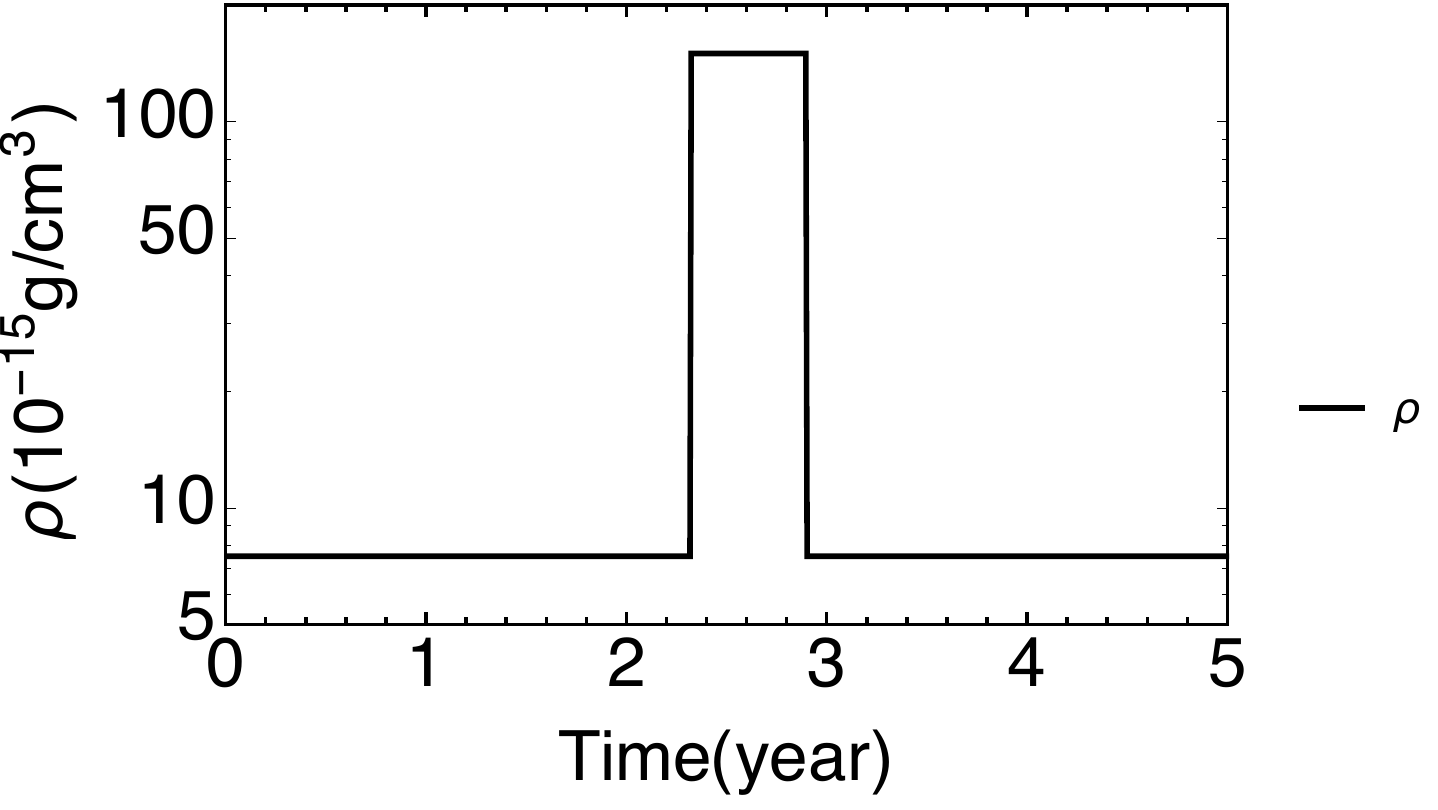}
    \includegraphics[scale=0.45]{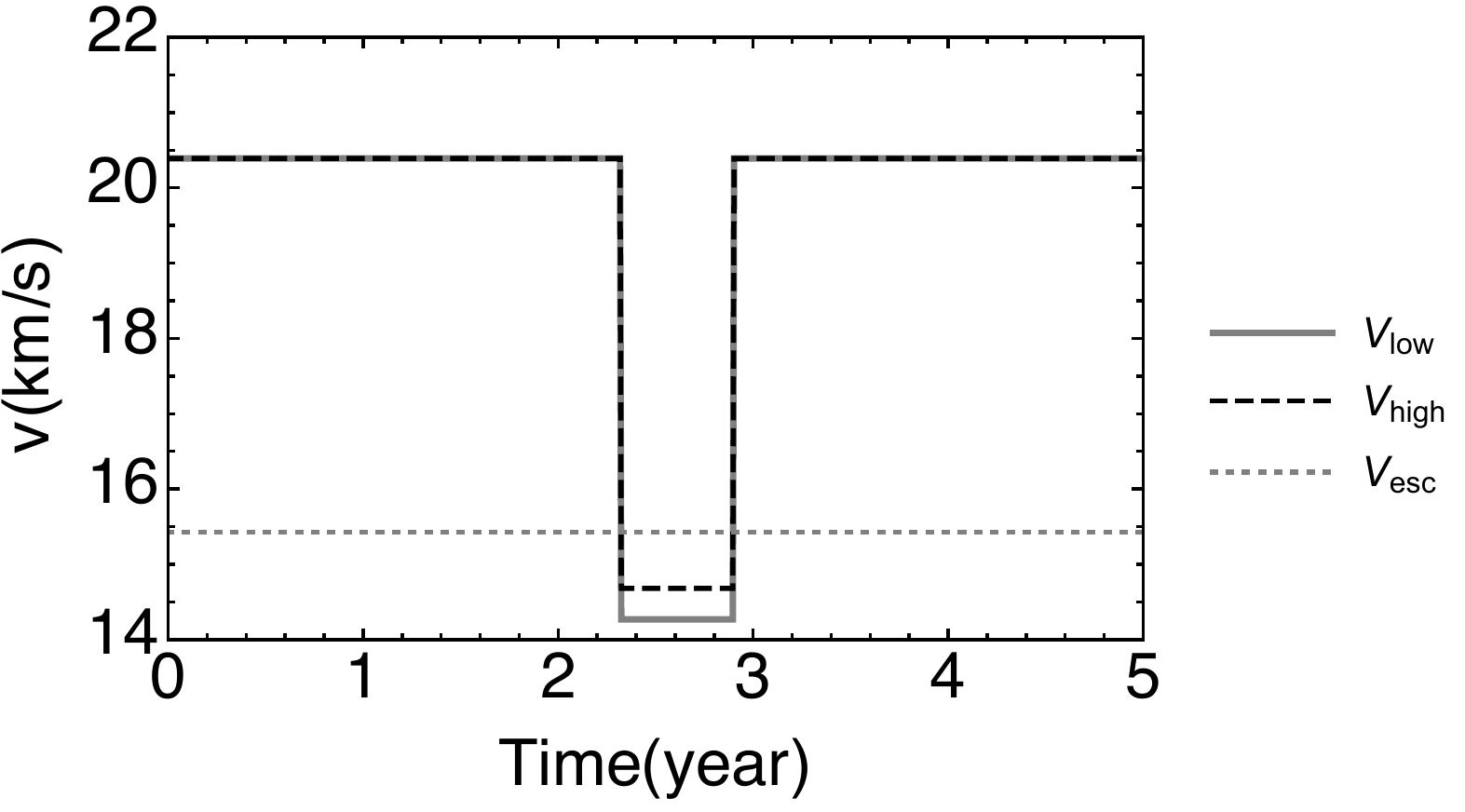}
    \caption{First panel shows the boundary condition of the star when it emits stellar wind. Second and third panels show time dependence of density and velocity.}
\end{figure}

We begin with an AGB star at the origin and allow it fill the grid with a pre-shell wind. The mass of the star is $1\ M_{\odot}$.  The escape velocity at $1\ AU$ from the center of the star is $v_{esc} = 15.43\ km/s$.  We account for radiation pressure within the gas and dust by using an the effective stellar gravity of $\alpha G M_{\odot}/r^2$ with $\alpha=0.134$. However, we do not specify the dust species and evolution in the simulation, that is, we only assume a steady state distribution of dust. The whole region can be optically thin which will be examined.
\begin{equation}\label{radiation}
f_{rad}=\frac{\rho(r)\kappa_{total}S(r)}{c}=\frac{(1-\alpha)G M_{\odot}\rho(r)}{r^2}
\end{equation}
therefore,
\begin{equation}\label{tau}
\tau=\int_{1AU}^{30AU}\rho(r)\kappa_{total}dr=\int_{1AU}^{30AU}\frac{4\pi(1-\alpha) G M\rho(r) c}{L}dr
\end{equation}
where $\rho(r)$ is the density of wind at radius of $r$, $\kappa_{total}$ is the total mass weighted opacity, $S(r)$ is the radiation intensity, $L$ is the bolometric luminosity of the AGB star and $\tau$ is the optical depth. In our simulation, the pre-shell wind is driven into the grid at $r_0=1\ AU$ with velocity  $v_0 = 20.39\ km/s$. The wind has an injection density of $\rho_0=7.5\times10^{-15}g/cm^3$ yielding a mass loss rate of $\dot{M}_0=6.84\times10^{-7}\ M_{\odot}yr^{-1}$. By mass conservation law,
\begin{equation}\label{density}
\rho_0 v_0 (AU)^2=\rho(r) v(r) r^2
\end{equation}
we can substitute equation (\ref{density}) into equation (\ref{tau}) and for computational convenience, eliminate $v_0$ and $v(r)$ on both side since they are on the same order of magnitude. For a typical $L=1500\ L_{\odot}$ AGB star, we can get that $\tau<1$. Obviously, if we lower all the wind density (as well as the mass loss rate) by a factor of several, the whole region would be optically thin and the hydrodynamic property would not change since equations (\ref{mass eq},\ref{momentum eq},\ref{idealgas}) all scale linearly with the density.

The simulation is isothermal and the temperature is $100\ K$. This yields an initial Mach number of $Mach= 22.4$. The boundary of the star is a sphere and the inner condition do not change upon the variation of outer condition. We divide the inner region of the star into two parts, a solid sphere and a spherically concentric shell (Shown in the first panel in Figure1.). The solid sphere has time dependent but uniform density and zero velocity while the concentric shell has the same time dependent uniform density but linearly increasing velocity from the inner side to the outer side. The inner side velocity is zero and the outer side velocity is the wind velocity. R and z axes are both reflective boundaries while top and right edge of the computational box only allow outflow, so any negative velocity at the boundary is set to be zero. The free fall time at the boundary is $80\ yrs$ and the sound speed is $0.90\ km/s$. Given that the initial ambient density is extremely low and that constant wind will fill the grid before the dense shell outflow, the boundary condition will have tiny effect on the simulation.

After the wind fills the grid we drive a dense shell from the star. The shell episode lasts from $t_1 = 2.32\ yr$ to $t_2 = 2.90\ yr$. During the episode, the density rises to $1.5\times10^{-13}\ g/cm^3$. The velocity drops to either $v_{low}=14.27\ km/s$ or $v_{high}=14.68\ km/s$ for the two cases we present here.  As we will see, these two cases bracket a bifurcation in the evolution of the ejected shell.  Both of these cases have $v_s < v_{esc}$  and in both cases approximately $M = 5.6\times 10^{-6}\ M_{\odot}$ is ejected in the shell. For computational convenience, we approximate the time dependence of density and velocity during shell ejections as step functions.

After the shell ejection terminates, we resume the steady wind (which we subsequently refer to as a "post-shell wind").  The second and the third panels in Figure1 show the density and velocity history for our simulations.  The figures show that in both simulations, the velocity during the shell ejection is sub-escape at the stellar boundary where mass flows into the grid.

Because \textsc{AstroBEAR} does not have 1-D capacities, we run our code in ($r,z$) cylindrical symmetry ($2.5D$) covering one quarter of the total plane and  average over the spherical angle $\theta$.  Thus we impose reflecting boundaries at $r=0$ and $z=0$. The top and right boundaries use outflow boundary conditions allowing material to leave the grid.  The winds and shell are injected into the grid via a spherical "surface" boundary condition at $r_0$.   The computational domain is $24\times24\ AU^2$ and consists of a $240\times 240$ base grids plus 4 levels of AMR. Thus the "effective" resolution of our simulations is $(3840)^2$.

\section{Results}
\subsection{Simulations from \textsc{AstroBEAR}}

In what follows we present  two examples from a larger suite of simulations chosen to bracket the transition from escape to fall-back modes of the ejected shell.  In Figure 2 and 3, we present $1D$ profiles of density and velocity at four different times in the evolution of the simulations.  We take the profiles along the 45-degree line in the $r-z$ plane.  We also plot the local escape velocity for reference.

\begin{figure}
    \centering
    \includegraphics[scale=0.35]{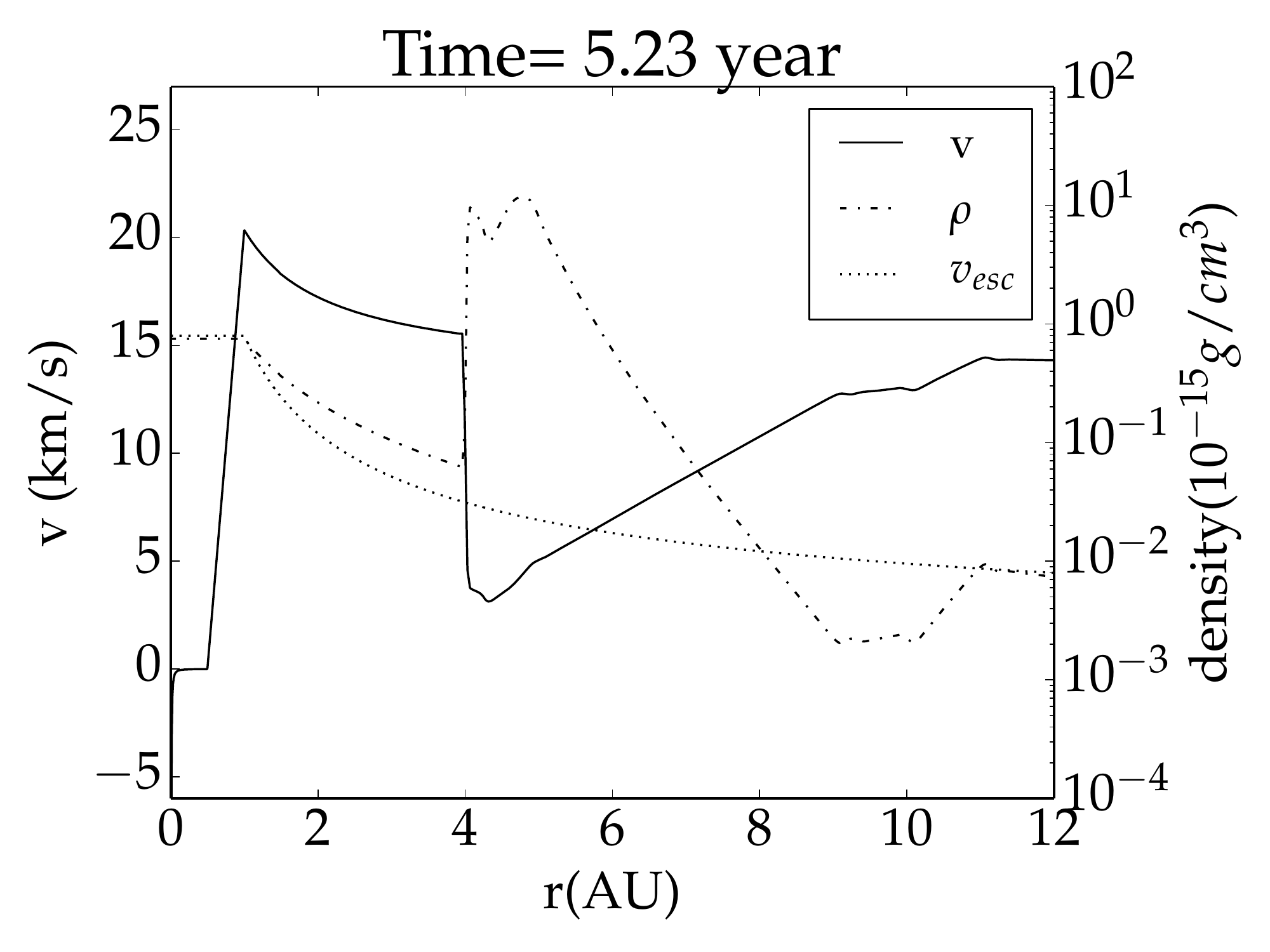}
    \includegraphics[scale=0.35]{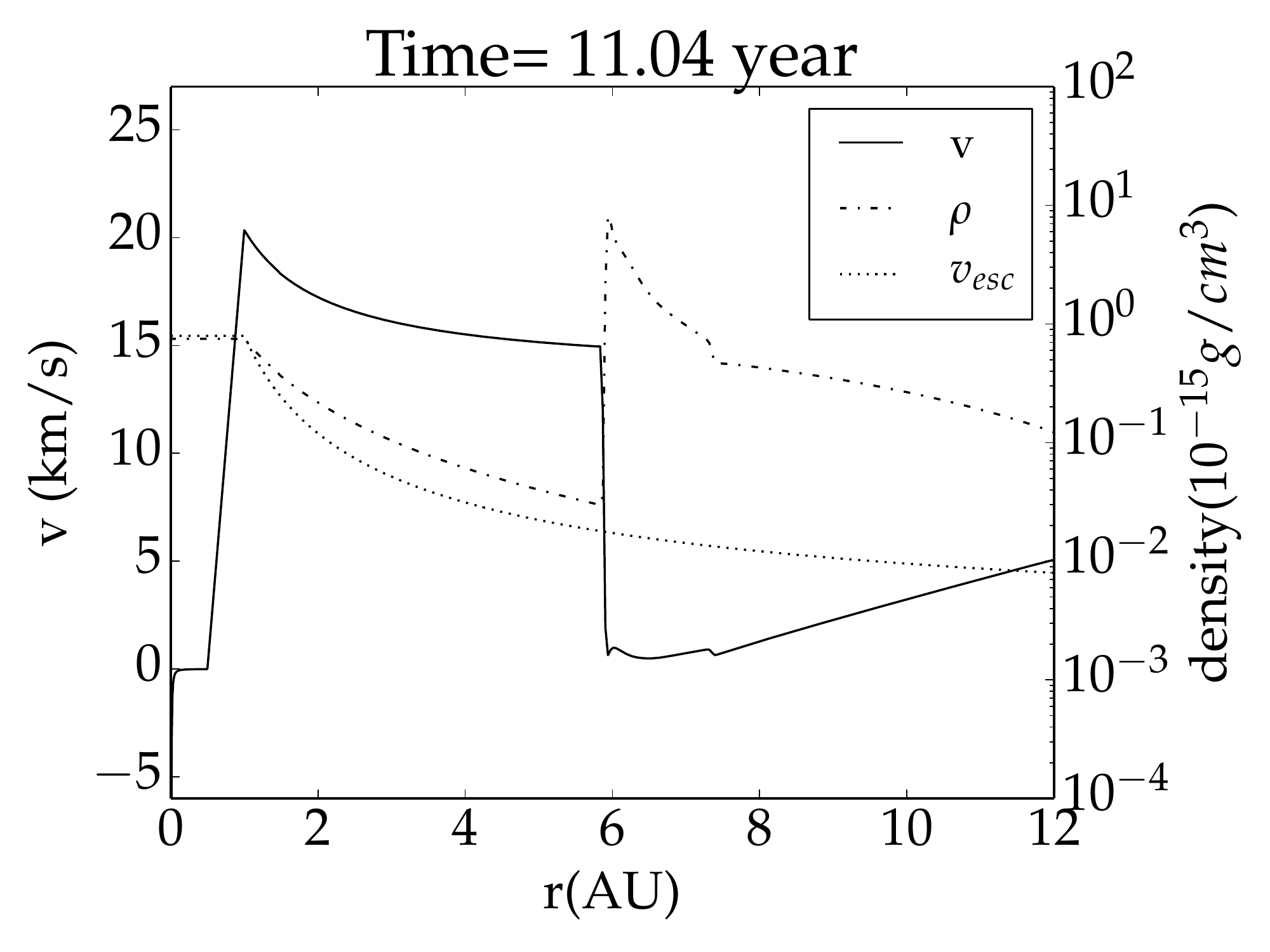}
    \includegraphics[scale=0.35]{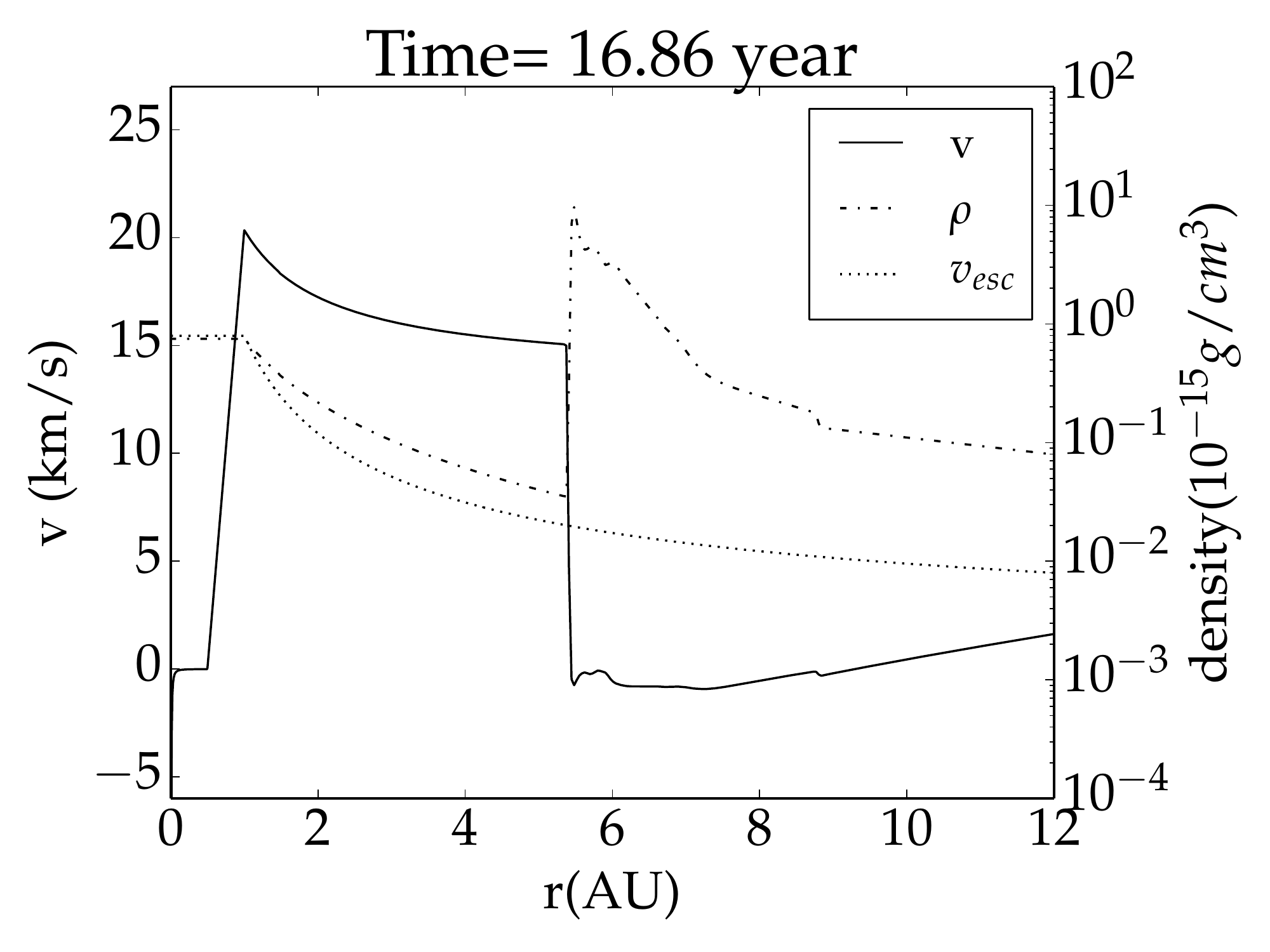}
    \includegraphics[scale=0.35]{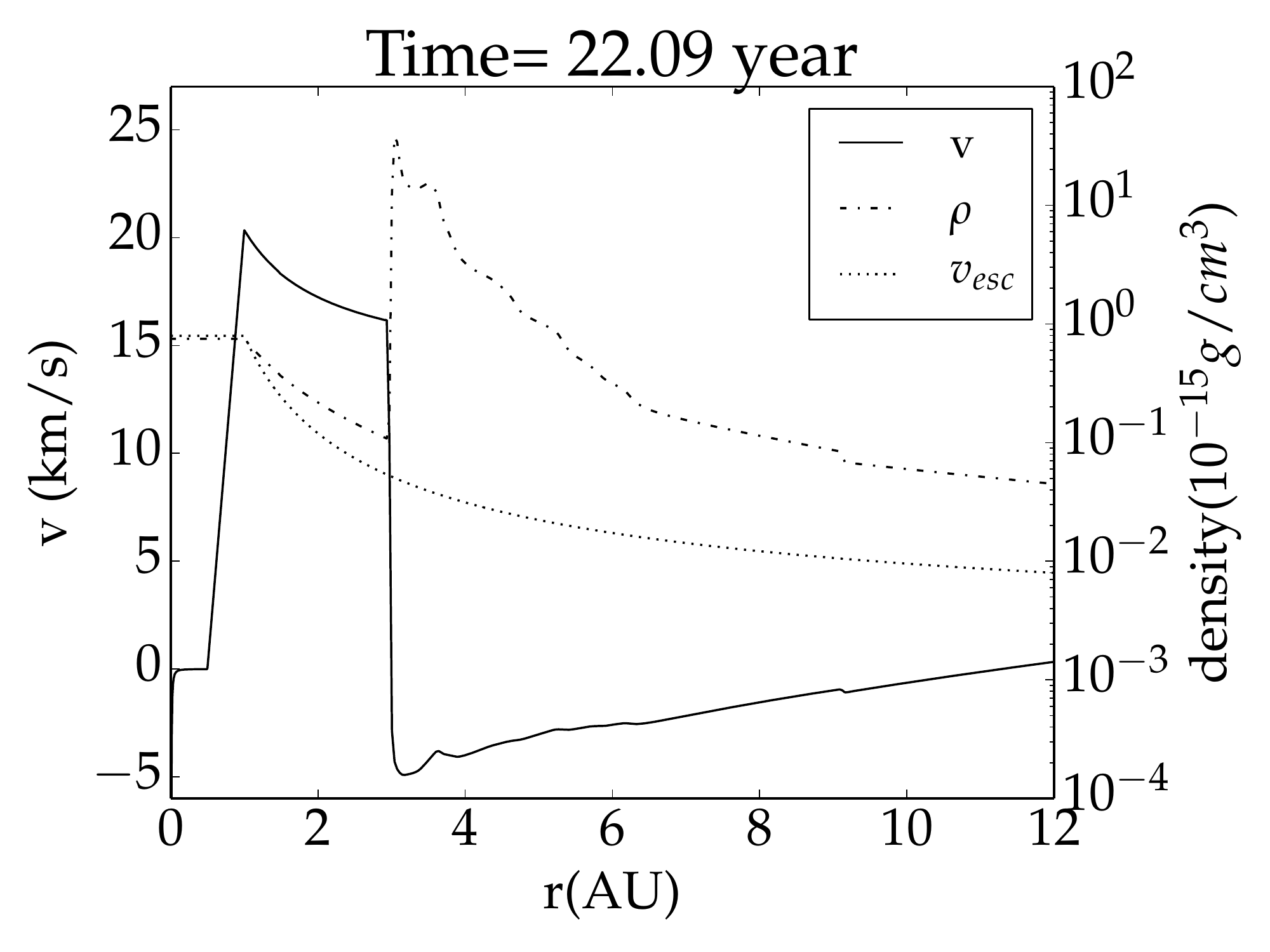}
    \caption{Time evolution of shell with initial shell velocity $v_{low}=14.27\ km/s$. The shell finally falls back.}
\end{figure}

The four panels in Figure2 show frames from the simulation with the lower shell ejection velocity $v_{shell}=14.27\ km/s$ (note with time increases from top to bottom).  The simulation begins with the steady wind filling the grid then the pulse of high density, sub-escape material (the shell) is launched at $2.32\ yr$.  In panel (a) we see the density and velocity profiles at at $t= 5.23\ yr$.  Note that the steep rise in velocity between the origin and $r= 1\ AU$ represents the inflow boundary where the stellar flow is injected into the grid.  In this region we also see that the density is held constant.

In the first panel between $1\ AU$ and $4\ AU$, the density of the steady wind ejected after the mass loss "pulse" drops continuously due to the geometric expansion of the gas.  The velocity also falls due to deceleration by the gravity of the AGB star. Thus since $\rho \propto \dot{M}_w/v_w(r) r^2$, the wind density in this region falls off more slowly that $\rho(r) \propto r^{-2}$.

At $r=4\ AU$, the steady post-pulse wind encounters the inner edge of the shell ejected during the brief pulse high mass loss.  From this shock interface to approximately $r= 6\ AU$ we see shell material.  Note that the velocity in this region is below the local escape velocity of the AGB star.    Between $r= 6\ AU$ and $r=9\ AU$ we see a steady drop in the density and steady increase in the velocity.  This region represents a rarefaction wave running from the denser but slower moving shell material and the supersonic, supra-escape pre-shell wind ejected before the pulse.  The most important point to note in panel (a) of figure 2 is the presence of the shock at $r=4\ AU$.  This shock wave represents a transfer of momentum from the post-pulse wind into the shell which  causes  varying degrees of shell acceleration depending, on the relative momenta of the wind and shell.

At $t=11.04\ yr$, we see the dense shell is has expanded outward.  The shock bounding the shell and the post-shell wind is now at $r=6\ AU$.  The boundary between the shell and pre-shell wind has expanded off the grid.  Note however that the velocity of the shell is just above $v=0$.  Thus the shell has been almost entirely decelerated and is reaching its maximum radial excursion.

By $t=16.86\ yr$, the velocity of the shell has fallen below $v=0$. At this point the shell has begun to fall back towards the star. By $t=22.09\ yr$, the shock boundary between the wind and the shell is at $r= 3\ AU$ as the shell continues to collapse back towards the star.  Note that for all positions above this radius we see $v<0$ which implies all material at these larger radii is moving radially inward as well.  This makes sense because of the rarefaction between the shell material and the pre-pulse wind. Since the right and top boundary is made to only allow outflow, it must be different from a self-consistent boundary. However, our simulation zone is large enough to keep the region we are interested unaffected. If the gas flows in from the boundary right after the velocity become zero, an error - induced by the artificial boundary condition - will travel at the sound speed which is $0.90\ km/s$. However, the error does not have enough time to travel from the boundary ($24\ AU$) to $12\ AU$ in just $22\ yrs$. So the fluid motion in $12\times12\ AU^2$ is not adversely affected by the boundary.

Finally, note that the sound speed becomes greater than shell speed as it decelerates ($cs > v$).  During this period we have a subsonic flow ($Mach<1$) where pressure force become important and we see an expansion of the shell width with very slow bulk radial motion.
  
\begin{figure}
    \centering
    \includegraphics[scale=0.35]{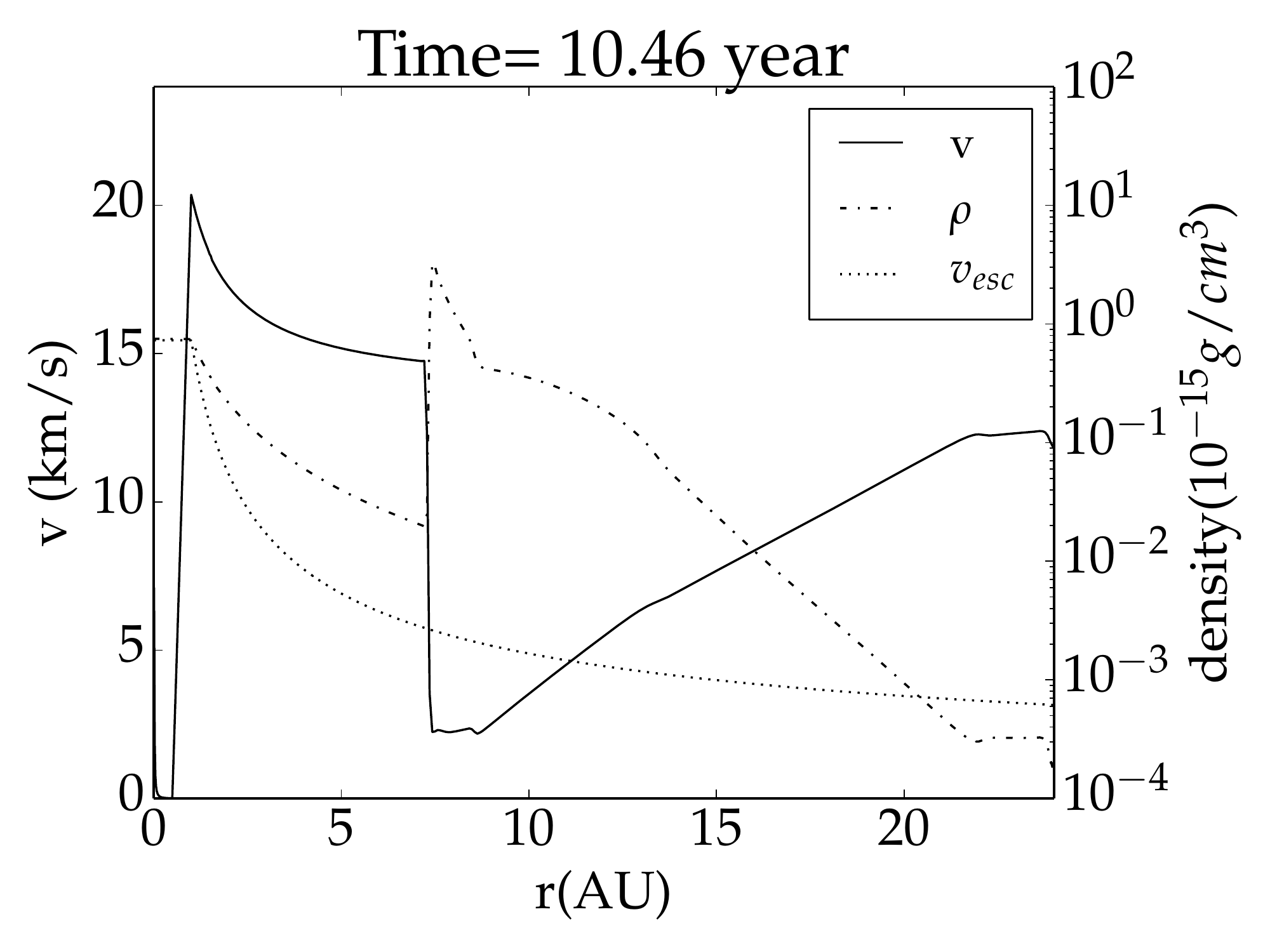}
    \includegraphics[scale=0.35]{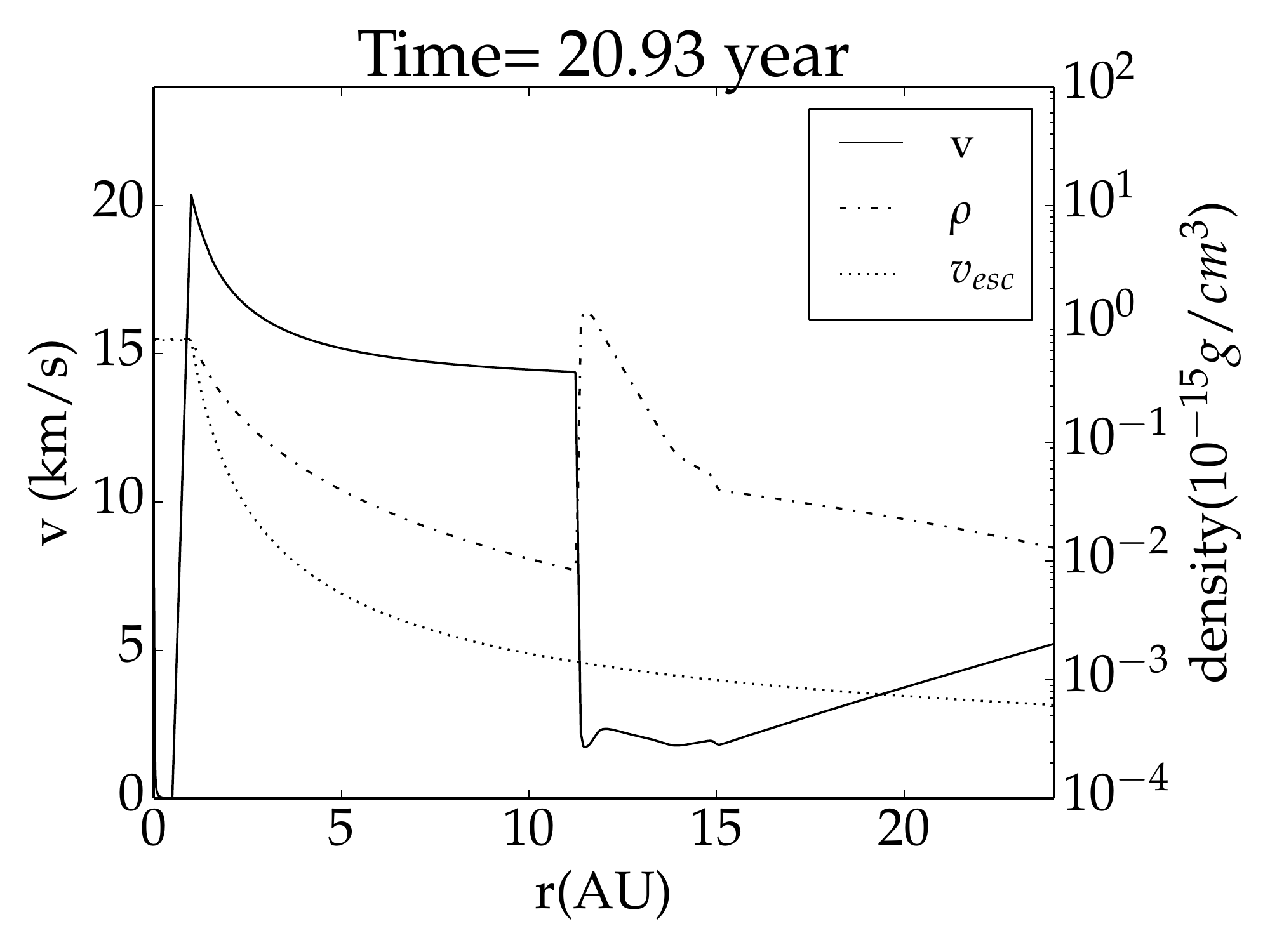}
    \includegraphics[scale=0.35]{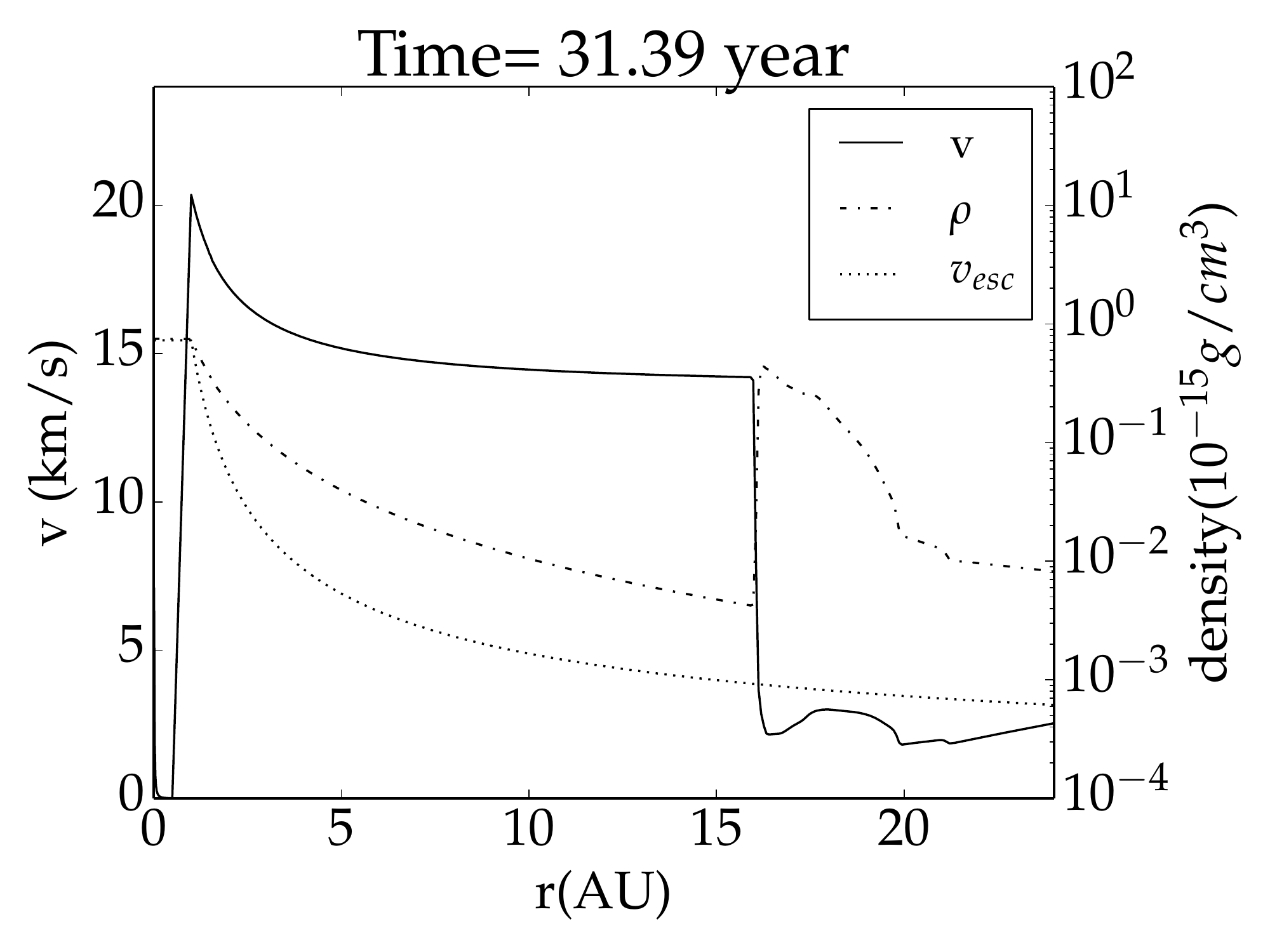}
    \includegraphics[scale=0.35]{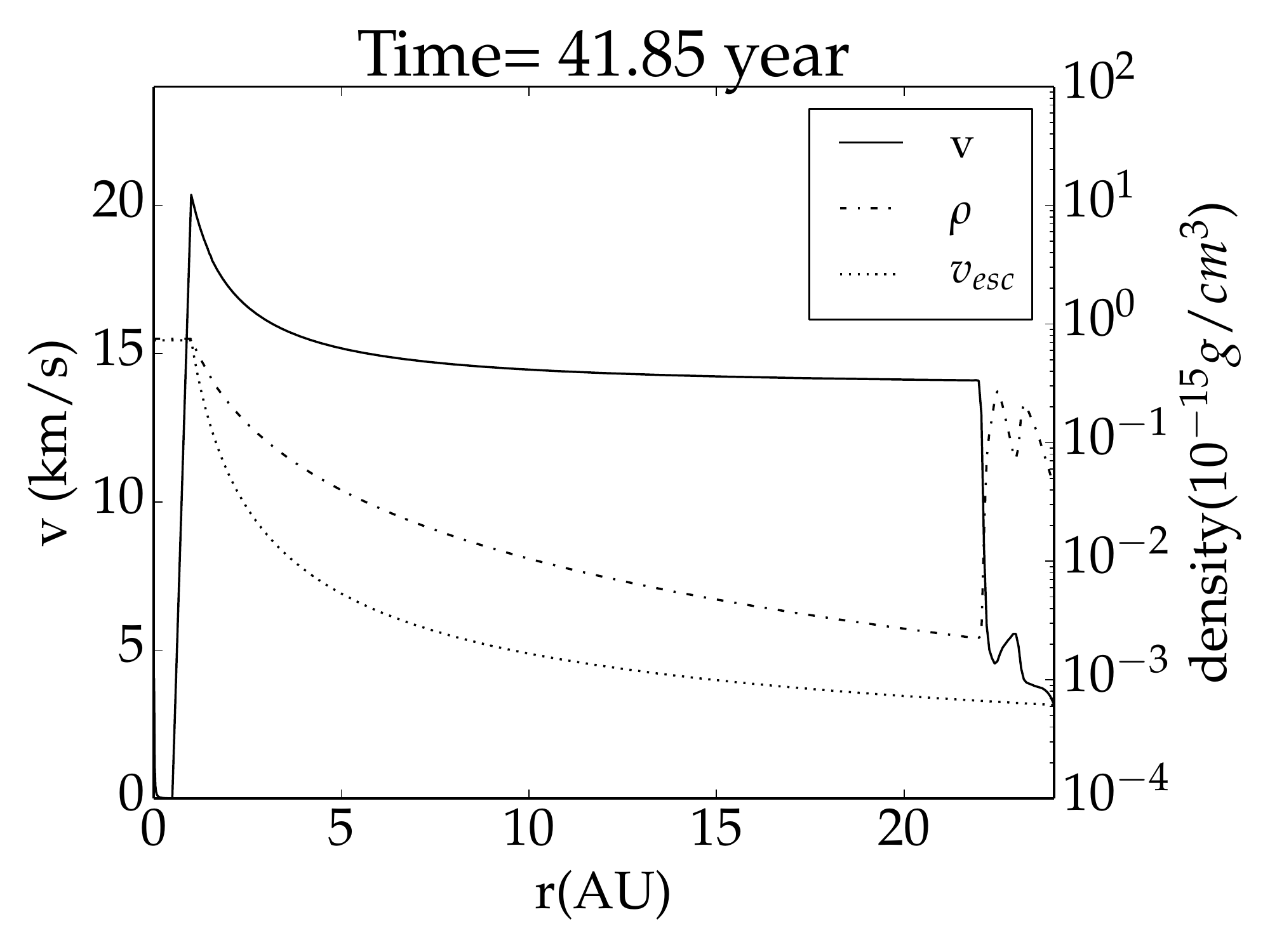}
    \caption{Time evolution of shell with initial shell velocity $v_{high}=14.68\ km/s$. The shell exceeds the escape velocity}
\end{figure}

In Figure 3, we show radial profiles of density and velocity for the simulation with a higher initial shell velocity.  In the first panel, taken at $t=10.46\ yr$, the shock boundary between post-shell wind and the shell is located at about $r=7.5\ AU$. Note the velocity of the shell is below the local escape velocity. As the post-shell wind continues to impart momentum to the shell, we observe a different evolution. In subsequent frames, the shell does not stall but is slowly driven outward. By $t=31.39\ yr$, the shell has reached $r=16\ AU$ and its velocity is now just below the local escape value.  By $t=41.85\ yr$ the shell has been almost entirely pushed off the grid and has achieved a speed in excess of the local escape velocity. Thus we find a shell ejected with $v_{shell}=14.27\ km/s$, will fall back onto the star.  The shell with $v_{shell}=14.68\ km/s$, which is also below the escape velocity, at the launch radius, will escapes.

\subsection{A One Dimensional Spherical Model}

In order to understand the transition between models with shells that escape under the action of the post-shell wind and those which fall back onto the star, we now derive a simplified analytical treatment of the problem.

To compare our $2.5D$ simulation with a $1D$ analytic model, we use fluid tracers in the simulations to track shell material.  We use tracer mass  weighted averages of the tracer's radial extent as a measure of shell's position vs time $r_s(t)$.
\begin{equation}
r_{s}=\frac{\int_{1AU}^{30AU}\rho_{tracer} r^3dr}{\int_{1AU}^{30AU}\rho_{tracer} r^2dr}
\end{equation}
We assume  that the initial velocity and density of the shell are $v_s$ and $\rho_s$ respectively. For the shell initial mass we have:
\begin{equation}\label{shell mass}
m_{s}=4 \pi r^2 \rho_{s} v_{s} \triangle t,
\end{equation}
where $\Delta t$ is the time interval of injection. The post-shell wind has initial velocity $v_0$ and initial density $\rho_0$ at the launch point.  For a steady spherical wind, mass conservation implies:
\begin{equation}
4 \pi r^2 \rho_w v_w=4 \pi r^2_0 \rho_0 v_0
\end{equation}
Thus the wind density will fall off as: $\rho_w=({r^2_0 \rho_0 v_0})/({r^2 v_w})$.

We calculate the evolution of the shell velocity $v_s$ from the momentum equation (\ref{momentum eq}). 
We assume that pressure forces are negligible with  $\|-\nabla P\| \ll \|\frac{G M \rho}{r^2}\|$ . This assumption is valid at all times except during the brief stalling of the shell that initiates fallback.  The momentum equation thus becomes
\begin{equation}\label{new momentum eq}
\frac{\partial \rho \mathbf{v}}{\partial t}+\nabla \cdot (\rho \mathbf{v} \mathbf{v})=-\frac{G M \rho}{r^2}\mathbf{\hat{r}}.
\end{equation}
We must track the velocity of the wind gas before it impacts the dense shell as it is subject to deceleration due to gravity, 
\begin{equation}\label{wind velocity}
v_{w}(r) = \sqrt{\mathbf{v}^2_0-\frac{2GM}{r_0}-\frac{2G M}{r}}
\end{equation}

For the momentum of the shell we write a discrete form of the momentum equation as
\begin{equation}
m_{s}(t+dt) \mathbf{v}_s(t+dt)=m_{s}(t) \mathbf{v}_{s}(t)+dm_{w} \mathbf{v}_{w}-\frac{G M m_{s}}{r^2}\mathbf{\hat{r}}dt
\end{equation}
where the second term on the right represents momentum added to the shell by the wind in a time $dt$ and the third term represents deceleration due to gravity.  More specifically,
for the wind mass added to the shell we have,
\begin{equation}
dm_{w}=4 \pi r^2 \rho (v_{w}(t)-v_{s}(t)) dt
\end{equation}
so that 
\begin{equation}
m_{s}(t+dt)=m_{s}(t)+dm_{w}
\end{equation}
Using the evolution of the shell velocity:
\begin{equation}\label{finaleq}
\mathbf{v_s}(t+dt)=\frac{m_{s}(t) \mathbf{v}_{s}+dm_{w} \mathbf{v}_{w}-\frac{G M m_{s}}{r^2}\mathbf{\hat{r}}dt}{m_{s}(t+dt)}
\end{equation}

\begin{figure}
        \centering
        \includegraphics[scale=0.5]{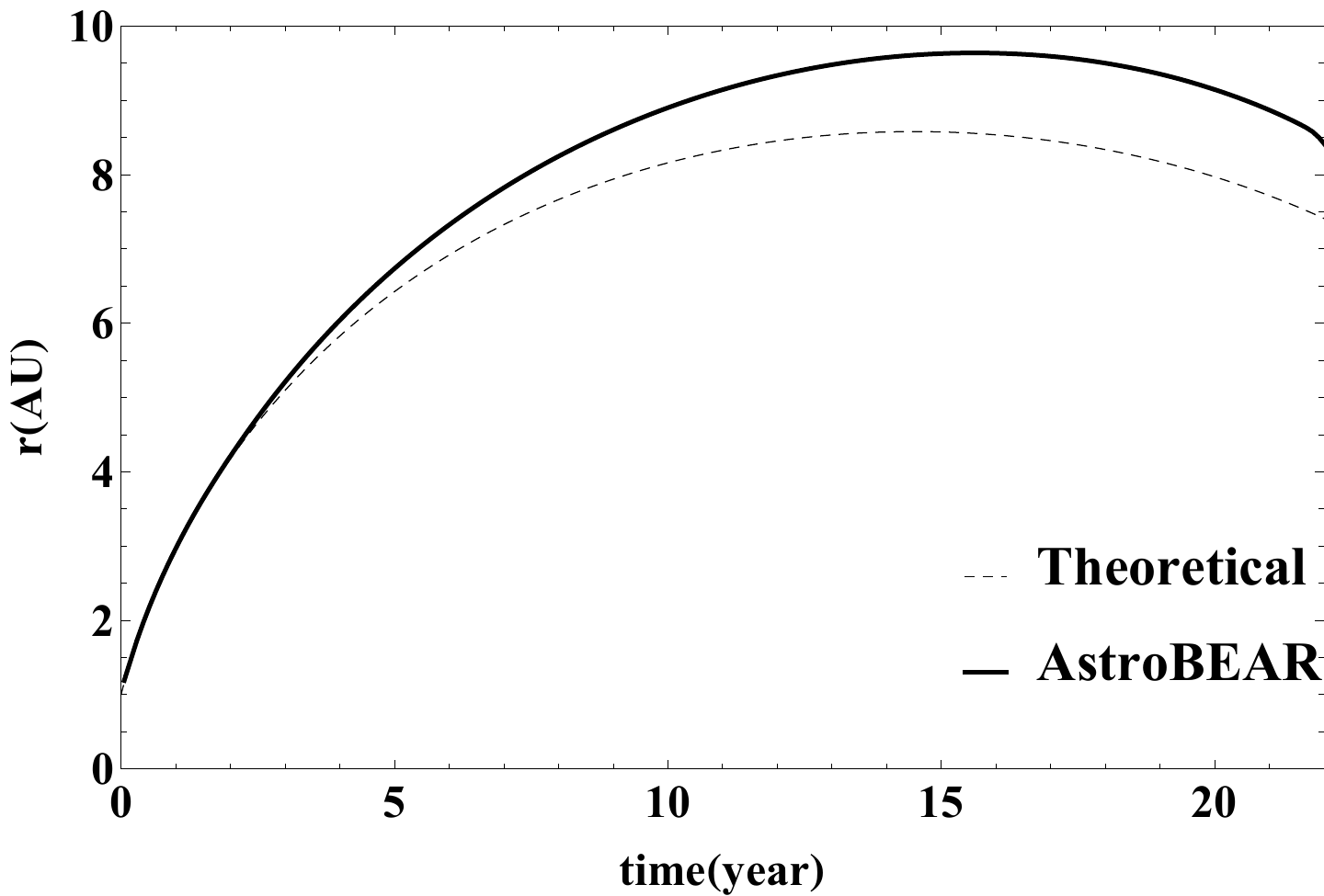}
        \includegraphics[scale=0.5]{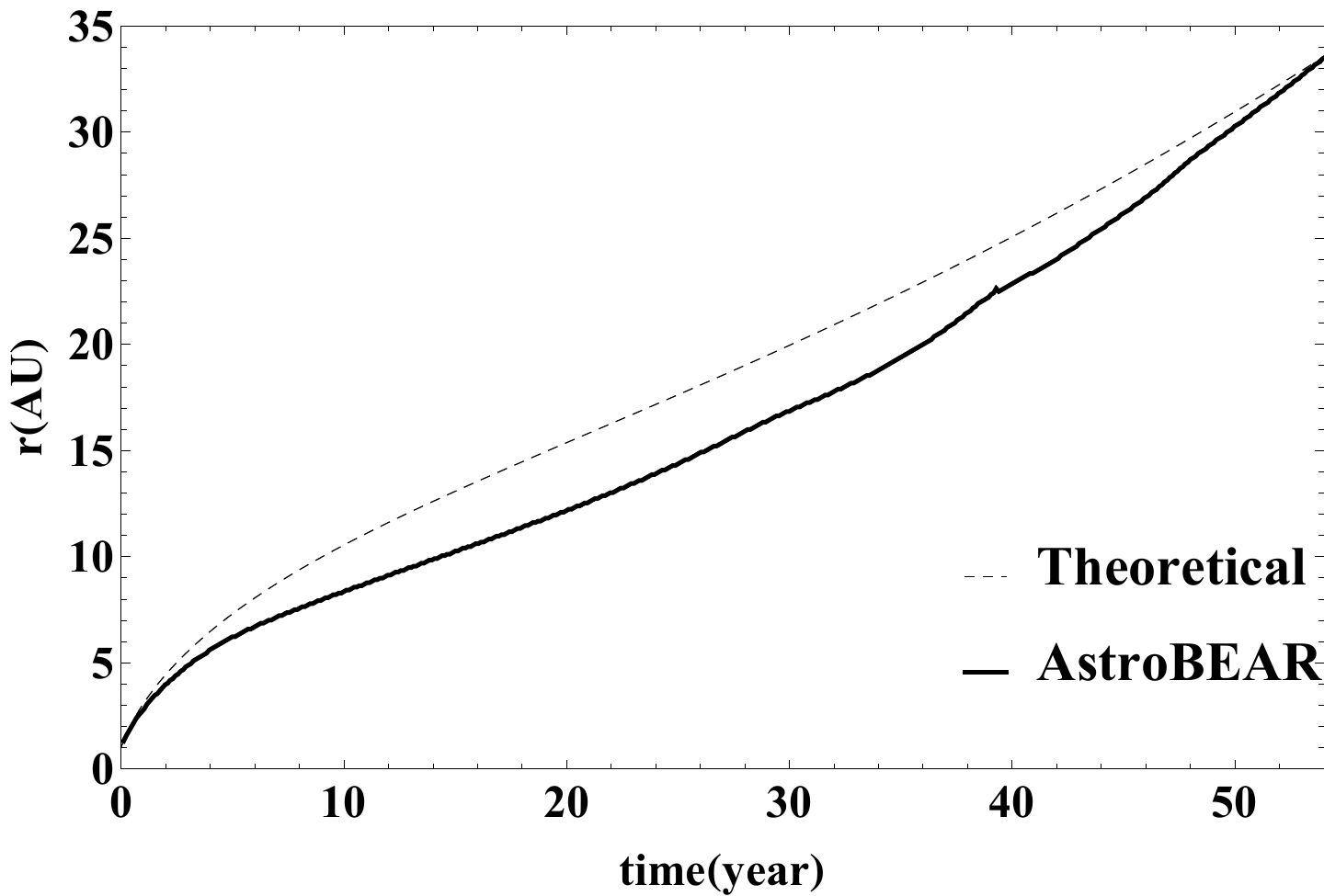}
        \caption{Results of trajectories of the shell from \textsc{AstroBEAR} and analytic model. Pulsation velocities are $v_{low}=14.27\ km/s$ and $v_{high}=14.68\ km/s$ respectively.}
\end{figure}

Using Eqs. (9) and (10), we divide equation (8) by $m_s(t+dt)$ and solve it numerically to find the trajectory of the shell $r(t)$ for any set of initial conditions.  In particular we look for the critical value of the initial shell velocity $v_c$ that distinguishes between fall back and escape.  Evaluation of equation (\ref{finaleq}) shows that for the inputs used in the simulations, the critical initial shell velocity is $v_c= 14.50\ km/s$.  Figure 4 shows a comparison of the $r_s(t)$ for the analytic model and for simulations (i.e. $v_0 < v_c$ and $v_0 > v_c$).  The analytic model does quite a good job of recovering the behavior seen in the simulations in both cases.

We can also estimate $v_c$ even more simply by first approximating the turning point of the shell in the fall back case.  Here assume that the mass of the shell is constant and neglect the effect of post-shell wind.  This gives us a ballistic approximation for the maximum radius of the shell $r_{m}$.  We then assume that the ram pressure of the steady wind must be large enough to support the shell against gravity at this radius if the wind is to stop the shell from falling back.  This yields
\begin{equation}\label{0th order criterion}
\frac{G m_{s} M}{r^2}=\frac{dp_{ram}}{dt}
\end{equation}
where $p_{ram}$ is the the magnitude of the ram momentum of the wind given by 
\begin{equation}
p_{ram} =4\pi \rho r_m^2 v_{w}^2=4\pi \rho_{0} r_0^2 v_0 v_{w}=\dot{M} v_{w}
\end{equation}
Note that $v_{w}$ is a function of $r$
from Eq. (\ref{wind velocity}).
  Note also that $r_m$ is determined by $v_{0}$ according to our assumption of ballistic motion, namely 
it corresponds to the radius where the velocity vanishes. Then force balance  gives 
\begin{equation}\label{0th order r}
r_m=\frac{-2 r_{0} GM}{v_{0}^2 r_0 - 2GM},
\end{equation}
\begin{equation}\label{0th order v}
v_{w}=\sqrt{v_c^2-\frac{2GM}{r_0}+\frac{2GM}{r_m}},
\end{equation}
Then using  Eqs. (\ref{0th order r}), (\ref{0th order v})
and \ref{shell mass} in the form 
\begin{equation}\label{0th order m}
m_{s}=4\pi r_{0}^{2} \rho_{0} v_c \triangle t
\end{equation}
in Eq. (\ref{0th order criterion}), we find  $v_c$. The  result is $v_c = 14.77\ km/s$ which is only 1.9\%  higher than the numerical result $14.50\ km/s$

\section{Summary and Discussion}

We have presented a simple model for the evolution of a shell ejected with sub-escape velocity during a brief pulse of enhanced mass loss in an AGB star.  The fate of the ejected shell depends on the acceleration to escape velocity via the action of the continuing steady AGB wind that follows the launching of the shell.  Our $2.5D$ isothermal hydrodynamic simulations are compared with a spherically-symmetric analytic model for the time evolution of the shell.  Both the analytic model and simulations closely support in their predictions for the value of the critical shell velocity $v_c$ that delineates the bifurcation between escape and fall-back modes of shell evolution. Though in the presence of spherically symmetric radial temperature gradient, the diffusion rate will be different but the qualitative result that we find--a bifurcation of fall back vs. escape modes would still be expected.

Our results lay out the basic dynamics of fall-back shells and are relevant to any mechanism which ejects material with less than the local escape velocity.  Our study also establishes the first step along the path to study post-AGB disk formation via fall-back shells.  Since AGB stars are slow rotators, we would expect that an ejected shell would not have enough angular momentum to establish a Keplerian disk at large radii. Thus ejected gas will either fall back onto the star or escape for isolated stars. In the case of a binary however, the material falling back onto the AGB star might gain enough angular momentum through gravitational interaction to form a Keplerian disk at large radii.  This is a  topic for future study.

\section*{Acknowledgments}

Financial support for this project was provided by the LLE at the University of Rochester, Space Telescope Science Institute grants HST-AR-11251.01-A, HST-AR-12128.01-A, HST AR-12146.04-A; by the National Science Foundation under awards AST-0807363 and AST-1102738; by the Department of Energy under award de-sc0001063.  EB acknowledges support from the Simons Foundation, and the IBM-Einstein fellowship fund at IAS.  The CIRC at the University of Rochester provided computational resources. A special thanks to Baowei Liu and Jonathan Carroll-Nellenback.

\bsp

\label{lastpage}

\end{document}